\documentclass[amsmath,amssymb,a4paper,onecolumn]{revtex4}

\usepackage{amssymb}
\usepackage{txfonts}
\usepackage{graphicx}
\usepackage{color}
\usepackage{array}

\begin{document}

\title{Berry Phase and Topological Effects of Phonons}

\author{Yizhou \surname{Liu}$^{1,2,3}$}
\author{Yong \surname{Xu}$^{1,2,3}$}
\email{yongxu@mail.tsinghua.edu.cn}
\author{Wenhui \surname{Duan}$^{1,2,4}$}
\email{dwh@phys.tsinghua.edu.cn}

\affiliation{
$^1$State Key Laboratory of Low Dimensional Quantum Physics, Department of Physics, Tsinghua University, Beijing 100084, People's Republic of China \\
$^2$Collaborative Innovation Center of Quantum Matter, Beijing 100084, People's Republic of China \\
$^3$RIKEN Center for Emergent Matter Science (CEMS), Wako, Saitama 351-0198, Japan \\
$^4$Institute for Advanced Study, Tsinghua University, Beijing 100084, People's Republic of China}

\maketitle

Phonons as collective excitations of lattice vibrations are the main heat carriers in solids. Tremendous effort has been devoted to investigate phonons and related properties, giving rise to an intriguing field of phononics, which is of great importance to many practical applications, including heat dissipation, thermal barrier coating, thermoelectrics and thermal control devices~\cite{li2012}. Meanwhile, the research of topology-related physics, awarded the 2016 Nobel Prize in Physics, has led to discoveries of various exotic quantum states of matter, including the quantum (anomalous/spin) Hall [Q(A/S)H] effects, topological insulators/semimetals and topological superconductors~\cite{hasan2010,qi2011}. An emerging research field is to bring topological concepts for a new paradigm phononics---``topological phononics''~\cite{zhang2010,kane2014,liu2016}.  In this Perspective, we will briefly introduce this emerging field and discuss the use of novel quantum degrees of freedom like the Berry phase and topology for manipulating phonons in unprecedentedly new ways.

Berry phase and topology are well established concepts for electrons and have shown important consequences on electronic properties~\cite{xiao2010,hasan2010,qi2011}. When varying the wavevector $\mathbf{k}$ along a closed loop within the Brillouin zone (BZ), the Bloch wavefunction  $\left| \psi_\mathbf{k} \right\rangle$ gains a geometric phase, called Berry phase, $\gamma = \oint\mathrm{d}\mathbf{k} \cdot \mathbf{A}_\mathbf{k} $, where $\mathbf{A}_\mathbf{k} = \left\langle \psi_\mathbf{k} \right| -i\nabla_\mathbf{k} \left| \psi_\mathbf{k} \right\rangle$ is the Berry connection. The Berry curvature $\mathbf{B}_\mathbf{k} = \nabla_\mathbf{k} \times \mathbf{A}_\mathbf{k}$ plays a role of magnetic field in the momentum space, which generates anomalous/spin Hall effects~\cite{xiao2010}. The integration of $\mathbf{B}_\mathbf{k}$ over the whole BZ gives a topological invariant, the Chern number $\mathcal{C}$, which characterizes the Q(A)H states~\cite{hasan2010,qi2011}.

One may generalize topological concepts from electrons to phonons straightforwardly by writing phonon equation in a Schr\"odinger-like form, $H_\mathbf{k} \psi_\mathbf{k} = \omega_\mathbf{k} \psi_\mathbf{k}$~\cite{liu2016}, where
\begin{equation}
H_{\mathbf k} = \left(
\begin{array}{ccc}
0                      & i D_{\mathbf k}^{1/2} \\
-i D_{\mathbf k}^{1/2}     & -2i{\eta}_{\mathbf k}
\end{array}
\right),~~~
{\mathbf \psi_{\mathbf k}} = \left(
\begin{array}{c}
  D_{\mathbf k}^{1/2} {\mathbf u_{\mathbf k}} \\
  {\dot{\mathbf{u}}_{\mathbf k}}
\end{array}\right).
\end{equation}
$D_{\mathbf k}$ is the dynamic matrix. ${\eta}_{\mathbf k}$ is the  time reversal symmetry (TRS) breaking term. $\mathbf u_{\mathbf k}$ and $\dot{\mathbf{u}}_{\mathbf k}$ are the displacement and its time derivative. $\omega_\mathbf{k}$ is the phonon dispersion. $\psi_\mathbf{k}$ is the phonon wavefunction expressed in an extended coordinate-velocity space. Similar equations with different definitions of wavefunction were proposed~\cite{zhang2010,susstrunk2016}. Based on the Schr\"odinger-like equation, the Berry connection, Berry curvature and other topological quantities are defined for phonons the same as for electrons.

However, the problem of phonons differs significantly from that of electrons on the following aspects~\cite{liu2016}. Firstly, phonons as bosons are not limited by the Pauli exclusion principle, suggesting that the whole phonon spectrum is physically probed. Secondly, spin-related effects (like QSH) cannot be directly applied to phonons that typically are spinless, but can be simulated by utilizing pseudo spin and pseudo spin-orbit coupling~\cite{susstrunk2015}. Last not least, while phonon Hamiltonians can be mapped to tight-binding electron Hamiltonians, they are required to satisfy the acoustic sum rule by the rigid translational symmetry. Moreover, their basis functions are always $p_{x,y,z}$, independent of the atomic nature, implying distinct orbital physics between phonons and electrons.

Honeycomb lattice is an ideal platform to explore Berry phase and topological effects of phonons~\cite{zhang2010,liu2016}. The longitudinal optical and acoustic modes form linearly crossed bands and Dirac points at $K$/$K^\prime$ (Fig. 1a), as protected by inversion symmetry (IS) and TRS. Phonons around the Dirac points are described by an effective Hamiltonian (referenced to the Dirac frequency)~\cite{liu2016}:
\begin{equation*}
H_0 = v_D(k_y\tau_z\sigma_x - k_x\sigma_y),
\end{equation*}
where $v_D$ is the group velocity, $\sigma$ and $\tau$ are the Pauli matrices, $\sigma_z=\pm1$ and $\tau_z=\pm1$ refer to $A$/$B$ sublattices and $K$/$K^\prime$ valleys, respectively. This Hamiltonian resembles that for spinless electrons of graphene. In the presence of IS and TRS, the Berry curvature is zero except at the Dirac points that are characterized by Berry phases of $\pm\pi$, leading to suppression of backscattering.

One may introduce mass terms into the Hamiltonian by breaking IS ($m_I$) or TRS ($m_T$) :
\begin{equation*}
H^\prime = m_I\sigma_z + m_T\sigma_z\tau_z,
\end{equation*}
which opens Dirac band gaps. The first term (the Semenoff term) can be realized by using different atomic masses for $A$/$B$ sublattices. The second term (the Haldane term) can be realized extrinsically by magnetic/Coriolis fields or intrinsically by spin-lattice interactions in magnetic lattices. Importantly, the effective Hamiltonian is based on $p_{x,y}$ orbitals rather than $p_z$, which is essential to get a nonzero Haldane term~\cite{liu2016}.

When slightly breaking TRS or IS, $\mathbf{B}_\mathbf{k}$ keeps vanishing except near $K$/$K^\prime$ (Figs. 1b,c). Then the Berry flux around each valley is $\pm\pi$, which gives the valley Chern number $\mathcal{C}_{K(K^\prime)} = \pm1/2$. By definition $\mathcal{C} = \mathcal{C}_{K}+\mathcal{C}_{K^\prime}$. If only TRS is broken, $\mathbf{B}_\mathbf{k} = \mathbf{B}_{-\mathbf{k}}$ and $\mathcal{C} = \pm1$, corresponding to the Q(A)H-like phase, which is topologically protected to have one-way edge phonon modes within the Dirac gap (Fig. 1b). These one-way edge modes are perfect transport channels, enabling conducting phonons without scattering. If only IS is broken, $\mathbf{B}_\mathbf{k} = -\mathbf{B}_{-\mathbf{k}}$ and $\mathcal{C} = 0$, corresponding to a topologically trivial phase. Nevertheless, for interfaces having opposite $m_I$ on the two sides, there exist topologically protected interfacial phonon modes that are valley-momentum locked (Fig. 1c). When both TRS and IS are broken, the frequency degeneracy between the two valleys is removed, and thus band gaps at $K$/$K^\prime$ can be tuned independently. One could realize zero band gap only at one valley in the extreme cases (Fig. 1d), which could be used for valley filters.

The Berry phase and topological effects of phonons can find promising applications. One possible direction is based on valley phononics that utilize the valley degree of freedom to control phonons. Valley phonons can be selectively excited by Raman processes using polarized photons~\cite{zhang2015}. Moreover, valley-polarized currents could be generated by the valley-polarized conduction channels (Fig. 1c) or by valley filters (Fig. 1d)~\cite{liu2016}, which can be further detected by valley phonon Hall effect~\cite{zhang2015}. Another possible direction is based on the one-way edge states or more generally the topologically protected boundary states. These states are able to conduct phonons with little or no scattering, useful for designing phononic circuits. Moreover, an ideal phonon diode with 100\% efficiency is possible in a multi-terminal transport system with one-way edge states~\cite{liu2016}.

As an outlook, many interesting research topics are awaiting to be explored in this field: (i) Find efficient ways to break TRS of phonons in solids. (ii) Search topological phonon states protected by symmetries like space group symmetries, with no need of breaking TRS. (iii) Explore Dirac or Weyl phonons in 3D lattices. (iv) Realizing the Berry phase and topological effects of phonons in realistic materials.

\begin{acknowledgements}
Y.X. acknowledges support from the National Thousand-Young-Talents Program and Tsinghua University Initiative Scientific Research Program. Y.L. and W.D. acknowledge support from the Ministry of Science and Technology of China (Grant No. 2016YFA0301001) and the National Natural Science Foundation of China (grant nos. 11674188 and 11334006).
\end{acknowledgements}

\begin{figure*}
\includegraphics[width=\linewidth]{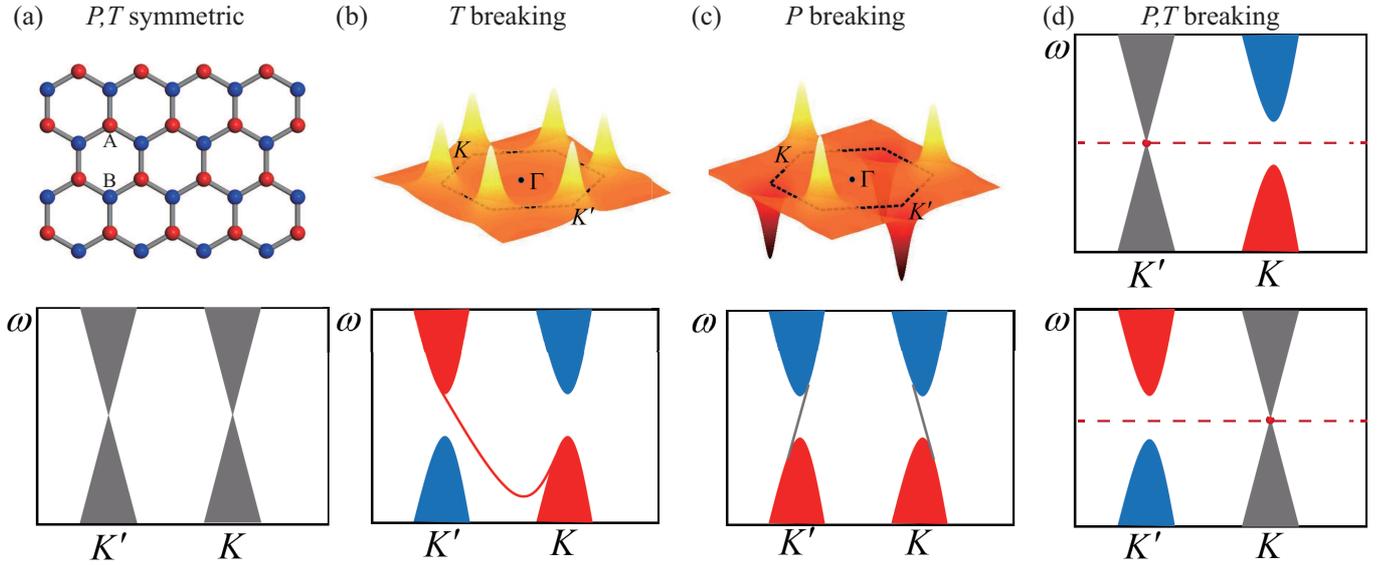}
\caption{\label{fig1} (a) A honeycomb lattice composed of $A$/$B$ sublattices, and the Dirac cones of phonons near the $K$/$K^\prime$ valleys. (b-c) Berry curvature distribution (upper), schematic bulk bands together with topological boundary states (lower) of phonons with broken (b) inversion symmetry $P$ and (c) time reversal symmetry $T$. Phonon states mainly located at $A$ ($B$) sublattices are colored red (blue). (d) The two valleys that are non-degenerate in frequency caused by breaking $P$ and $T$ simultaneously.}
\end{figure*}

\end{document}